\documentclass[floatfix,%
 reprint,
 amsmath,amssymb,
 aps,
]{revtex4-2}
\usepackage{float} 
\usepackage{graphicx}
\usepackage{booktabs}
\usepackage{dcolumn}
\usepackage{bm}

\usepackage{appendix}

\newcommand{\td}{\tilde{\delta}}
\newcommand{\mnras}{Monthly Notices of the Royal Astronomical Society}
\newcommand{\aap}{Astronomy and Astrophysics}
\newcommand{\jcap}{Journal of Cosmology and Astrophysics}

\newcommand{\physrep}{Physical Reports}
\newcommand{\pasp}{Publications of the Astronomical Society of the Pacific}
\usepackage{hyperref}

\begin{document}
\preprint{APS/123-QED}

\title{Neutrino Mass Signatures in the Galaxy Bispectrum}

\author{Farshad Kamalinejad}%
 \email{f.kamalinejad@ufl.edu}
\affiliation{%
Department of Physics, University of Florida, 2001 Museum Rd., Gainesville, FL 32611}%
\author{Zachary Slepian}
\email{zslepian@ufl.edu}
 \affiliation{Department of Astronomy, University of Florida, 211 Bryant Space Science Center, Gainesville, FL 32611}

\date{\today}

\begin{abstract}
In the Standard Model, neutrinos are massless. However, oscillation experiments demonstrate that they do have a small mass. Currently, only the differences of the masses squared are known, along with an upper bound on their sum. Upcoming surveys of the Universe's Large-Scale Structure (LSS) offer a promising avenue to probe neutrino mass by revealing how neutrinos influence galaxy clustering. Massive neutrinos affect mode coupling within the framework of Perturbation Theory (PT) for structure formation, leaving detectable signatures in the PT kernels. In this work, we present for the first time the explicit modifications to the kernels caused by massive neutrinos and investigate the extent of these changes in the redshift-space galaxy bispectrum. To this end, we generate synthetic data using a theoretical covariance matrix and employ Markov-Chain Monte Carlo (MCMC) to assess the impact of these new signature terms. This approach, in contrast to Fisher forecasting, allows us to see if the new terms induce shifts in the recovered central values of parameters. The synthetic data is produced to mirror two different galaxy samples. The first corresponds to the Sloan Digital Sky Survey (SDSS) Baryon Oscillation Spectroscopic Survey (BOSS) Data Release 12 CMASS Luminous Red Galaxy (LRG) sample, characterized by an effective volume of $V_{\rm eff} = 3 \;[{\rm Gpc}/h]^3$ and a number density of $\bar{n} = 3 \times 10^{-4} \;h/{\rm Mpc}$. The second corresponds to the Dark Energy Spectroscopic Instrument (DESI) Year 5 LRG sample with $V_{\rm eff} = 25 \;[{\rm Gpc}/h]^3$. Our findings indicate that neglecting neutrino mass effects in the kernels can result in a central value shift equivalent to approximately $1\sigma$ in the galaxy biases estimated from the DESI Y5-like sample. For the BOSS CMASS volume, the shift, though not statistically significant, is non-negligible.

\end{abstract}

\maketitle


\section{\label{sec:level1}Introduction}
With the recent discovery of the Higgs boson \citep{HiggsBoson2012}, the last missing piece of the Standard Model (SM)---or the first hint of beyond-SM physics---is the neutrino mass scale and hierarchy. Terrestrial experiments such as $\beta$-decay \citep{katrin2001katrin} struggle to reveal either: cosmology offers a nearly unique avenue forward. According to neutrino oscillation experiments \citep{Fukuda_1999, Ahmad_2002,Eguchi_2003} neutrinos are massive with three neutrino mass flavors. Our current paradigm predicts a relic neutrino background with number density of $112\; {\rm cm}^{-3}$ per neutrino flavor.  

Due to the mass, this relic neutrino background must become non-relativistic at some point in the Universe's history, as its kinetic energy is diluted by cosmic expansion, and so eventually becomes subdominant to the rest-mass energy. The time of this transition is set by the mass, and in turn imprints a scale on the clustering of matter, potentially enabling large-scale structure (LSS) surveys to detect $\sum m_{\nu}$ , the sum of the three mass flavors \citep{Wong_2011, Basse_2014, Audren_2013}. 

Neutrino oscillation experiments also reveal that at least two of the neutrino mass states are non-relativistic at present. The tightest bound on the neutrino mass to date is given by combining Cosmic Microwave Background (CMB) and Baryon Acoustic Oscillation (BAO) measurements. By doing so, the Dark Energy Spectroscopic Instrument (DESI) collaboration finds a 95\% confidence upper bound of $\sum m_{\nu}<0.072\;{\rm eV}$ \citep{adame2024desi}, which is much tighter than the CMB-only measurement from the \textit{Planck} satellite, $\sum m_{\nu}<0.26\;{\rm eV}$ \citep{aghanim2020planck}.

Given the tremendous volume of data to become available in the next decade via efforts such as DESI (2019-2025) \citep{DESI:2016} and Euclid (2022-2028) \citep{scaramella2022euclid}, a robust detection of $\sum m_{\nu}$ should be within the reach of near-term cosmology. However, extracting it with certainty from galaxy surveys is not trivial. The typical approach is to measure a decrement in the galaxy power spectrum, which is the Fourier transform (FT) of the two-point correlation function (2PCF). The 2PCF measures the excess of galaxy pairs as a function of their separation, relative to the expectation from a spatially random distribution. This decrement is studied on scales within the horizon that neutrinos reach once they become non-relativistic. The neutrinos are homogeneous on scales smaller than this and so suppress density fluctuations on such scales \citep{Hu_1998, LESGOURGUES_2006, Levi2016_neutrino, Saito_2008, Wong_2008}. 

The decrement in the power spectrum cannot be measured with high precision for several reasons. First, we do not observe the matter field directly, but rather the galaxies, which are a biased tracer of the underlying matter field (\citep{Desjacques:2018} for a review). The galaxy biases (and potentially the Effective Field Theory of Large-Scale Structure (EFT) counter-terms \citep{Senatore:2015, matarrese1997large, angulo2015statistics, de2020impact}) could mimic this decrement, making the neutrino mass degenerate with these biases in practice. Furthermore, the biases themselves are degenerate with the amplitude of the power spectrum (encoded by $\sigma_8$, the root mean square (RMS) of the density fluctuations on $8\;{\rm Mpc}/h$ spheres). As a result, the neutrino mass becomes degenerate with $\sigma_8$ in the power spectrum \citep{hahn2020constraining, Hahn_2021, Yankelevich_2018}.

Other than the galaxy power spectrum, the galaxy bispectrum (FT of the triplet correlation function, or 3PCF, \textit{e.g.} \citep{SE_3PCF_BAO, SE_3PCF_model, SEcomp}) is sensitive to the neutrino mass (\citep{Scoccimarro1999, Bernardeau_2002} for an important initial paper, and a review, covering the the bispectrum). The galaxy bispectrum has already proven a powerful probe of the cosmological parameters \citep{yankelevich2023halo, Yankelevich_2018, hahn2020constraining, Ruggeri_2018}, capable of breaking degeneracies present in both the galaxy power spectrum and the CMB \citep{philcox2022boss, ivanov2023cosmology}; it is also a key tool to explore Primordial Non-Gaussanity (PNG)  \citep{sefusatti2006cosmology, scoccimarro1998nonlinear, scoccimarro2000bispectrum, baldauf2015bispectrum, dizgah2021primordial, fergusson2009shape,baldauf2011primordial,liguori2010primordial}.

The bispectrum, unlike the power spectrum, arises directly from coupling of different Fourier modes of the density field, known as ``mode coupling". This coupling is described by mode-coupling kernels (or simply kernels). While the lowest order power spectrum does not depend on mode-coupling kernels, the bispectrum, at lowest order, consists only of a \textit{tree-level} contribution that is directly linked to the kernels and therefore, will be affected by any modifications of them.

Neutrinos also affect how different Fourier-space modes couple together. In this paper we show that this results in the emergence of a signature term in the Standard Perturbation Theory (SPT) kernels (reviewed in \citep{Wong_2008}). These kernels describe how higher-order density and velocity perturbations evolve from the linear-order terms. Consequently, neutrinos not only leave a measurable effect on the power spectrum (and subsequently on the bispectrum) due to small-scale suppression, but they also modify nonlinear structure formation, which will impact the higher-order (nonlinear) corrections to the power spectrum, and the tree-level bispectrum. 

Previously,  \citep{aviles2021clustering, Aviles_2020, Wong_2008} calculated the kernels numerically and obtained their effects on the redshift-space and real-space power spectra. In this paper, we calculate the neutrino imprint on the kernels, offer an analytical correction to the SPT kernels, and explore the extent of their impact on the redshift-space bispectrum. 

We also explore the potential impact of these new signature terms on parameter inference. To do this, we generate synthetic data and examine the posterior distribution derived from it. The synthetic dataset is constructed using a theoretical covariance matrix \citep{biagetti2022covariance}. We analyze this dataset through Markov Chain Monte Carlo (MCMC) analysis \citep{noriega2024unveiling}. The results of this analysis show whether  the signature term in the kernels is truly observable in the redshift-space galaxy bispectrum or if we can safely use the ``no-neutrino mass'' (standard SPT) kernels for future datasets.

This paper is structured as follows. In \S\ref{sec:PT}, we review SPT, discuss the free-streaming physics, and derive the neutrino-included Euler equations. Then, in \S\ref{sec:level2}, we obtain the neutrino signature kernels. In \S\ref{sec:level3}, we compute the redshift-space bispectrum monopole (with respect to the line of sight), including the neutrino signature kernels.  \S\ref{sec:synthetic} discusses how the synthetic data is created and what we use for the bispectrum covariance matrix. Finally, in  \S\ref{sec:Bis_Param_post}, we present the corner (triangle) plots obtained from the MCMC analysis and discuss the effects of the signature terms. The appendix structure is as follows. In \S\ref{sec:MCMCvsFisher}, we compare MCMC with the Fisher forecast and further justify this method. Additionally, \S\ref{sec:Binning} discusses the reasoning behind our choice of binning.

\section{Perturbation Theory Bispectrum }
\label{sec:PT}
Cosmological PT expresses the late-time matter density field as, essentially, a Taylor series around the initial conditions. Each successive term corresponds to a higher power of the linear density fluctuation field, $\delta_1 (\bf x) \equiv \rho_{\rm lin}(\bf x)/\bar{\rho}_{\rm lin} - 1$, with $\rho_{\rm lin}(\bf x)$ the linear-theory density, $\bar{\rho}_{\rm lin}$ the average density, and $\delta_1 (\bf x) \ll 1$ for PT to be valid \citep{Bernardeau_2002}. 

Typically we work in Fourier space such that any density (linear or nonlinear) is expanded as:
\begin{align}
\delta(\mathbf{x},\tau)=\int_{\mathbf{k}}e^{i\mathbf{k}\cdot \mathbf{x}} \,\td(\mathbf{k},\tau) 
\end{align}
and 
\begin{align}
    \td(\mathbf{k},\tau)=\int e^{-i\mathbf{k}\cdot \mathbf{x}}\, \delta(\mathbf{x},\tau) d^3{\mathbf{x}}
\end{align}
where $\int_{\mathbf{k}} \equiv \int {d^3\mathbf{k}}/{(2\pi)^3}$. 

In an Einstein-de Sitter (EdS) universe, we may expand the matter density fluctuation $\td_{\rm m}$ in a series ordered by powers of the scale factor $a(\tau)$, with $\tau$ conformal time. This approach is known as Eulerian Standard Perturbation Theory (SPT), and we have (\textit{e.g.} \citep{Bernardeau_2002}):
\begin{align}
    \td_{\rm m}(\mathbf{k},\tau) &= \sum_{n=1}^\infty a^n(\tau)\,\td_n(\mathbf{k});
    \label{eqn:ansatz}
\end{align}
the FT of the matter velocity divergence, $\tilde{\theta}_{\rm m}(\mathbf{k}, \tau)$, can be expanded similarly. We note that now $\td$ on the right-hand side is time independent, as the time-dependence is captured through the powers of $a$.

The matter is treated as a fluid (though see \citep{Carrasco_2012}), and solving the perturbed fluid equations (continuity, Euler, and Poisson) using the ansatz Eq. (\ref{eqn:ansatz}) results in higher-order density and velocity divergence terms as:
\begin{align}
    \td_n(\mathbf{k})&=\int_{\mathbf{q}_1}\cdots\int_{\mathbf{q}_n}\delta^{[3]}_{\rm D}\left(\mathbf{k}-\sum_{i=1}^n \mathbf{q}_i\right) F_n(\mathbf{q}_1,\ldots,\mathbf{q}_n)\nonumber\\
    &\times\td_1(\mathbf{q}_1)\cdots\td_1(\mathbf{q}_n)\nonumber\\
    \tilde{\theta}_n(\mathbf{k})&=\int_{\mathbf{q}_1}\cdots\int_{\mathbf{q}_n}\delta^{[3]}_{\rm D}\left(\mathbf{k}-\sum_{i=1}^n \mathbf{q}_i\right)\ G_n(\mathbf{q}_1,\ldots,\mathbf{q}_n)\nonumber\\
    &\times\td_1(\mathbf{q}_1)\cdots\td_1(\mathbf{q}_n),
\end{align}
where $F_n$ and $G_n$ are, respectively, the SPT density and velocity kernels, and $\delta_{\rm D}^{[3]}$ is a 3D Dirac delta distribution which enforces momentum conservation. 

To initialize the recursion giving $F_n$ and $G_n$, one sets $F_1 = G_1 = 1$; at $n=2$ we then have 
\begin{align}
&F_2(\mathbf{q}_1,\mathbf{q}_2)=\frac{17}{21}\mathcal{L}_0(\hat{q}_1 \cdot \hat{q}_2) +\frac{1}{2}\left(\frac{q_1}{q_2}+\frac{q_2}{q_1} \right)\mathcal{L}_1(\hat{q}_1 \cdot \hat{q}_2)\nonumber\\
&\qquad \qquad \qquad \; +\frac{4}{21}\mathcal{L}_2(\hat{q}_1 \cdot \hat{q}_2),\label{Eq:F2}\\
&G_2(\mathbf{q}_1,\mathbf{q}_2)=\frac{13}{21}\mathcal{L}_0(\hat{q}_1 \cdot \hat{q}_2) +\frac{1}{2}\left(\frac{q_1}{q_2}+\frac{q_2}{q_1} \right)\mathcal{L}_1(\hat{q}_1 \cdot \hat{q}_2)\nonumber\\
&\qquad \qquad \qquad \; +\frac{8}{21}\mathcal{L}_2(\hat{q}_1 \cdot \hat{q}_2),\label{Eq:G2}
\end{align}
where $\mathcal{L}_{\ell}$ is the order-$\ell$ Legendre polynomial. 

The leading-order (tree-level) matter bispectrum represents correlations of matter density fluctuations at three different wave-vectors, as \citep{Scoccimarro1999}:
\begin{align}
    B_{\rm m}(\mathbf{k}_1,\mathbf{k}_2,\mathbf{k}_3)\,(2\pi)^3&\delta_{\rm D}^{[3]}\left(\sum_{i=1}^3 \mathbf{k}_i\right) \\\nonumber&=\langle \td_{\rm m}(\mathbf{k}_1)\td_{\rm m}(\mathbf{k}_2)\td_{\rm m}(\mathbf{k}_3)\rangle
    \label{Eq:Bis_def}
\end{align}
where $\langle \rangle$ means taking an expectation value over many possible draws from the distribution  of the initial density fluctuations. According to the ergodic hypothesis, we take this as equivalent to averaging over many different spatial regions of the Universe.

We note that PT does not predict the location of each galaxy, but rather the statistical properties of the density field. Therefore, all the information is encoded in the $N$-Point Correlation Functions (NPCFs) of the density field (though on very small scales see \citep{Carron_2012}). At linear order, since the density field is believed to be a Gaussian Random Field (GRF), all the odd-point correlation functions vanish and even-point functions can be expressed in terms of the linear power spectrum $P(k)$ via Wick's Theorem \citep{wick1950evaluation}. The power spectrum is defined via \citep{Bernardeau_2002}
\begin{align}
   P(k)=\langle \td_1(\mathbf{k}_1) \td_1(\mathbf{k}_2) \rangle (2\pi)^3 \delta_{\rm D}^{[3]}(\mathbf{k}_1+\mathbf{k}_2).
\end{align}

However, since gravitational structure formation induces non-linearity the matter bispectrum offers additional information. Yet \textit{galaxy} redshift surveys observe not the matter bispectrum but the galaxy bispectrum. Galaxies are biased tracers of the matter, where the bias coefficients encode unknown details of galaxy formation physics over which we often marginalize when fitting models. 

We adopt a bias scheme as in  \citep{Desjacques:2018}, to second order in the matter density field \citep{fry1993biasing, Desjacques:2018}:
\begin{align}
    \td_{\rm{g}} = b_1 \td _{\rm{m}} + \frac{b_2}{2}\td_{\rm m}^2+b_{\mathcal{G}_2} \mathcal{G}_{2}.
    \label{eqn:bias}
\end{align}

$\td_{\rm g}$ is the Fourier-space \textit{galaxy} density fluctuation, $b_1$ the linear bias, $b_2$ the nonlinear bias, $b_{\mathcal{G}_2}$ the tidal tensor bias, and $\mathcal{G}_{2}$ the matter's tidal tensor. We adopt this bias model here because the three biases above are the only ones for which there is robust observational evidence in bispectrum or 3PCF measurements (\citep{Gil_Marin_2015}, \citep{SE_3PCF_BAO}). We neglect stochastic contributions \citep{Desjacques:2018}. 

Using Eq. (\ref{eqn:bias}) we may compute the real-space galaxy bispectrum as 
\begin{align}
\label{eqn:gal_bispec}
     &B_{\rm g}(\mathbf{k}_1,\mathbf{k}_2,\mathbf{k}_3)=2b_1^3 P(k_1)P(k_2)\\ &\times\left[F_2(\mathbf{k}_1,\mathbf{k}_2) + \gamma \mathcal{L}_0(\hat{k}_1 \cdot \hat{k}_2) +  \frac{2\gamma'}{3} \mathcal{L}_{2}(\hat{k}_1 \cdot \hat{k}_2)\right] + {\rm cyc.}\nonumber
\end{align}
where $\gamma \equiv b_2/b_1$ and $\gamma' \equiv b_{\rm t}/b_1$; $``\rm cyc."$ means to add the two additional permutations corresponding to the explicitly-given right-hand side term but evaluated at $(\mathbf{k}_2, \mathbf{k}_3)$ and $(\mathbf{k}_3, \mathbf{k}_1)$.

\subsection{Perturbation Theory with Massive Neutrinos}
\label{Sec:Perturbation_Theory_with_nu}
Neutrinos with mass $m_{\nu,i}$ ($i=1, 2, 3$ indexes the neutrino mass states) became non-relativistic at a redshift of $1+z_{{\rm nr},i} \approx 1894 \;(m_{\nu,i}/{1\; \rm eV})$ \citep{lesgourgues_mangano_miele_pastor_2013}, which means they started contributing to the matter energy density at that time \citep{Levi2016_neutrino}. Based on the most recent data from oscillation experiments, at least two of the neutrino flavors must be non-relativistic at present. In LSS cosmology, neutrinos are typically treated as if the flavors have equal masses since the primary observable is the sum of their masses. Therefore, in this paper, we consider three degenerate neutrino mass states. 

Now, the neutrino energy density $\Omega_{\nu}$, and the mass fraction, $f_{\nu}$ are defined as \citep{Blas_2014}:
\begin{align}
    \Omega_{\nu}=\frac{\sum m_{\nu}}{93.14\; h^2\; {\rm eV}},\;\;\;f_{\nu}=\frac{\Omega_{\nu}}{\Omega_{\rm m}},
\end{align}
where $\Omega_{\nu}$ and $\Omega_{\rm m}$ are the neutrino and matter energy densities in units of the critical density. $h$ is the Hubble constant in units of $100\;{\rm km}\;{\rm s}^{-1}\;{\rm Mpc}^{-1}$. Throughout this work, we explicitly state the time-dependence of the parameters when intended, and if this is not done, we are referring to their present-time values. For instance, $\Omega_{\rm m}(\tau)$ is time-dependent, whereas $\Omega_{\rm m}$ means the matter density parameter at present.

Since the neutrinos are free-streaming \citep{LESGOURGUES_2006, Levi2016_neutrino, kamalinejad2022two}, the length scale corresponding to their sound speed is the maximum distance they can propagate within a Hubble time. This length scale is known as the \textit{free-streaming} scale (in configuration space), and its inverse is $k_{\rm FS}$, the free-streaming wave-number (in Fourier space). In the limit where $z \ll z_{\rm nr}$, when the neutrino mass fraction $f_{\nu}$ becomes constant \citep{lesgourgues_mangano_miele_pastor_2013, Hu_1998, kamalinejad2022two}:
\begin{align}
    k_{\rm FS}(\tau) &= \sqrt{\frac{3\Omega_{\rm m}(\tau)\mathcal{H}^2(
    \tau)}{2c_s^2(\tau)}}\nonumber\\&\sim 1.035 \Omega_{\rm m}^{1/2} \left(\frac{\sum m_{\nu}}{3} \right) a(\tau)\;[h\rm{\,Mpc^{-1}}].
    \label{Eq:kF}
\end{align}
where $\mathcal{H}(\tau)$ denotes the Hubble parameter in terms of the conformal time, and $c_s(\tau)$ represents the neutrino sound speed.

The free-streaming scale has two important limits: first, its value at the transition redshift, which we call the \textit{non-relativistic} scale, $k_{\rm NR}$, and second, its value at $z=0$, $k_{\rm FS}(a=1)\equiv k_0$. {\bf The most interesting behavior of the power spectrum occurs between these two scales, where we observe a gradual suppression from $k_{\rm NR}$ to $k_0$ \citep{kamalinejad2022two}.} On scales larger than the non-relativistic scale $(k<k_{\rm NR})$, neutrinos do not affect the matter power spectrum, while on smaller scales than $k_0$ $(k>k_0)$, they cause a constant suppression of approximately $(P-P_{f_{\nu}=0})/P \sim - 8 f_{\nu}$ \citep{Hu_1998}. In the case of the bispectrum, this suppression is more significant, with $(B-B_{f_{\nu}=0})/B \sim -13.5 f_{\nu}$ \citep{Levi2016_neutrino}.

We are setting out to calculate the effects of neutrinos on the PT kernels based on the approach of  \citep{Wong_2008}. In their treatment, they retain the neutrino density contrast at linear order but neglect the neutrino velocity perturbations. Additionally, in this approach, the free-streaming scale $k_{\rm FS}$ does not directly enter the equations, as we have manually set the velocity perturbations of neutrinos to zero (as we are considering regimes where $k\gg k_{\rm FS}$). Consequently, this approach is known to violate momentum conservation \citep{Blas_2014}. 

However, \citep{Blas_2014} finds that this approximation only impacts the 2-loop and higher corrections to the power spectrum. Therefore, for the tree-level redshift-space bispectrum, which only depends on the $F_2$ and $G_2$ kernels (as we will see later), the approximation is valid. Using this also provides an analytical framework for studying massive neutrino effects on the kernels. 

\citep{Wong_2008} shows that the fluid equations for the cold dark matter (CDM) + baryons become:
\begin{align}
    &\partial_s \Psi_i(\mathbf{k},s) + K_{ij}\Psi_j(\mathbf{k},s)+N_{ij}\Psi_j^{\nu}(\mathbf{k},s) = \nonumber\\&\int d^3\mathbf{q}_1 d^3\mathbf{q}_1\,\delta_{\rm D}^{[3]}(\mathbf{k}-\mathbf{q}_{12})\gamma_{ijl}\Psi_j(\mathbf{q}_1,s) \Psi_l(\mathbf{q}_2,s)
    \label{One_fluid_eq}
\end{align}
where $s\equiv\ln{a}$ denotes the time dependence. The indices \( i \) and \( j \) refer to the \( i \)-th (or \( j \)-th) element of a vector or the \( (i,j) \)-th element of a matrix, respectively.
$\Psi(\mathbf{k},s)$ is the 2D vector of CDM+baryons density and velocity perturbations, and $\Psi^{\nu}(\mathbf{k},s)$ the same for neutrinos:
\begin{align}
    \Psi(\mathbf{k}, s) = \begin{bmatrix}
\delta_{\rm cb} \\ -1/\mathcal{H}\;\theta_{\rm cb}
\end{bmatrix},\,\,\,
    \Psi^{\nu}(\mathbf{k},s) = \begin{bmatrix}
\delta_{\nu} \\ 0
\end{bmatrix}.
\end{align}
The term in $K_{ij}$ captures the homogeneous part of the two-fluid equations, and $N_{ij}$ the inhomogeneous part. The $K_{ij}$ are elements of the matrix given by:
\begin{align}
    K &= \begin{bmatrix}
0 & -1 \\
-\frac{3}{2}(1-f_{\nu}) & \frac{1}{2}
\end{bmatrix}
\end{align}
and the $N_{ij}$ are elements of the matrix given by
\begin{align}
    N &= \begin{bmatrix}
0 & 0 \\
-\frac{3}{2}f_{\nu} & 0
\end{bmatrix}
\end{align}
$\gamma_{abc}$ (in Eq. (\ref{One_fluid_eq})) is the mode-coupling kernel and is zero except for $\gamma_{121} = \alpha(\mathbf{q}_1, \mathbf{q}_2)$ and $\gamma_{222} = \beta(\mathbf{q}_1, \mathbf{q}_2)$. We remind the reader that the mode-coupling kernels $\alpha$ and $\beta$ act only on CDM+baryons, and have no effect on neutrino density perturbations as a result of our approximation. Therefore, the kernels derived in this paper should apply only to the CDM+baryons contribution to the total matter density, and not to the neutrinos, since the neutrinos do not cluster on the scales that we are exploring.

%
\section{\label{sec:level2}Neutrino Signatures}
As noted earlier, neutrinos cause a constant suppression on small scales in both the matter power spectrum and bispectrum. {\bf It is worth considering whether the effect of neutrinos on the PT kernels offers a more distinctive, scale-dependent signature.} So motivated, we now outline how neutrinos modify the SPT kernels to leading order in $f_{\nu}$. 

We denote the modified kernels by $\mathcal{F}$ and $\mathcal{G}$. Following \citep{Wong_2008}, our ultimate goal here is to derive simple corrections to the bispectrum that are valid at order $f_{\nu}$. Solving Eq. (\ref{One_fluid_eq}) order by order yields the updated kernels \citep{Wong_2008}:
\begin{align}
\label{eqn:script_kernels}
&\mathcal{F}_2(\mathbf{q}_1,\mathbf{q}_2) = \frac{15A_1 + 2A_4}{21} \mathcal{L}_0(\hat{q_1} \cdot \hat{q}_2) \\
&+ \frac{1}{2}\left( A_2\frac{q_1}{q_2}+ A_3\frac{q_2}{q_1}\right)\mathcal{L}_1(\hat{q}_1 \cdot \hat{q}_2) + \frac{4A_4 }{21}\mathcal{L}_2(\hat{q}_1 \cdot \hat{q}_2),\nonumber\\
&\mathcal{G}_2(\mathbf{q}_1,\mathbf{q}_2) = \frac{9C_1 + 4C_4}{21} \mathcal{L}_0(\hat{q_1} \cdot \hat{q}_2) \\
&+ \frac{1}{2}\left( C_2\frac{q_1}{q_2}+C_3\frac{q_2}{q_1}\right)\mathcal{L}_1(\hat{q}_1 \cdot \hat{q}_2) + \frac{8 C_4 }{21}\mathcal{L}_2(\hat{q}_1 \cdot \hat{q}_2).\nonumber
\end{align}
The $A$ and $C$ coefficients are:
\begin{align}
    A_1&=\frac{7}{10}\sigma_{11}^{(2)}(q_1,q_2)[f(q_1)+f(q_2)],\\
    A_2&=f(q_2)[\sigma_{11}^{(2)}(q_1,q_2)+\sigma_{12}^{(2)}(q_1,q_2)f(q_2)],\\
    A_3&=f(q_1)[\sigma_{11}^{(2)}(q_1,q_2)+\sigma_{12}^{(2)}(q_1,q_2)f(q_1)],\\
    A_4&=\frac{7}{2}\sigma_{12}^{(2)}(q_1,q_2)f(q_1)f(q_2),\\
    C_1&=\frac{7}{6}\sigma_{21}^{(2)}(q_1,q_2)[f(q_1)+f(q_2)],\\
    C_2&=f(q_2)[\sigma_{11}^{(2)}(q_1,q_2)+\sigma_{12}^{(2)}(q_1,q_2)f(q_2)],\\
    C_3&=f(q_1)[\sigma_{11}^{(2)}(q_1,q_2)+\sigma_{12}^{(2)}(q_1,q_2)f(q_1)],\\
    C_4&=\frac{7}{4}\sigma_{22}^{(2)}(q_1,q_2)f(q_1)f(q_2),
\end{align}
where  $f(q)\equiv \partial \ln{D(q,\tau)}/\partial \tau$ is the logarithmic derivative of the linear growth rate $D(q,\tau)$, and we have suppressed its dependence on $\tau$. We also note here that the growth rate can be scale-dependent, due to the neutrinos \citep{Wong_2008, Hu_1998}. 

From the equations above, we see that $A_2$ and $A_3$ are not symmetric in their arguments considering all powers of $f_{\nu}$ ($A_2(q_1, q_2) \neq A_2(q_2, q_1)$, and the same for $A_3$) yet up to the linear-order Taylor expansion in $f_{\nu}$, they will be symmetric. It is also apparent that $A_2(q_1, q_2) = A_3(q_2, q_1)$. ${\bf \sigma}^{(2)}$ is a $2 \times 2$ matrix:
\begin{align}
\sigma^{(2)}(q_1,q_2) &\equiv \frac{1}{\mathcal{N}^{(2)}}
\begin{bmatrix}
    2\omega^{(2)}+1 \hspace{15pt} & 2 \\[10pt] 
    3(1-f_{\nu}) \hspace{15pt} & 2\omega^{(2)}
\end{bmatrix}, 
\end{align}
and 
\begin{align}
    &\omega^{(2)}(q_1,q_2)\equiv f(q_1)+f(q_2),\nonumber\\
    &\mathcal{N}^{(2)}\equiv(2\omega^{(2)}+3)(\omega^{(2)}-1)+3f_{\nu}. \nonumber
\end{align}
Given the growth rate \( D(q,\tau) \), we first compute \( f \), from which we then determine \( \mathcal{N}^{(2)} \) and \( \omega^{(2)} \), and subsequently the matrix \( {\bf \sigma}^{(2)} \). This, in turn, enables the explicit calculation of the coefficients \( A_i \) and \( C_i \), and ultimately of \( \mathcal{F}_2 \) and \( \mathcal{G}_2 \).

Along the same lines as our earlier point regarding the free-streaming scale (paragraph below Eq. \ref{Eq:kF}), most of the impact from the neutrino mass can be found on scales smaller than $k_{0}$. Since the neutrino fraction $f_\nu \ll 1$, we expand $f(q,\tau)$ to $\mathcal{O}(f_{{\nu}})$ in the regime where $k \gg k_{\rm NR}$ as \citep{Hu_1998}:
\begin{align}
    f(k,\tau)=\left(1-\frac{3}{5}f_{\nu}\right)f_{\nu=0}(\tau),
    \label{eqn:bare_D}
\end{align}
where $f_{\nu=0}(\tau)$ is the logarithmic growth rate in an EdS cosmology in the absence of neutrinos. 

The restriction to $k\gg k_{\rm NR}$ is not a severe one in practice. It is unlikely that we can ever measure the bispectrum on scales $k<k_{\rm NR}$ precisely enough to add much to the neutrino mass constraint, given the limited cosmic volume available to us and the concomitantly high cosmic variance on large scales. 

Following the steps outlined above in this section we can obtain the modified kernels as:
\begin{align}
    \boxed{
    \begin{aligned}
        \mathcal{F}_2(\mathbf{q}_1,\mathbf{q}_2) &= F_2(\mathbf{q}_1,\mathbf{q}_2) \\
        &\quad + \frac{4}{245} \big[ \mathcal{L}_0(\hat{q}_1 \cdot \hat{q}_1) 
        - \mathcal{L}_2(\hat{q}_1 \cdot \hat{q}_1) \big] f_{\nu}
    \end{aligned}
    }
    \label{F2kernel}
\end{align}
and

\begin{align}
    \boxed{
    \begin{aligned}
        \mathcal{G}_2(\mathbf{q}_1,\mathbf{q}_2) &= G_2(\mathbf{q}_1,\mathbf{q}_2)
        - \bigg[
        \frac{83}{245} \mathcal{L}_0(\hat{q}_1 \cdot \hat{q}_1) \\
        &\quad + \frac{3}{10}\left(\frac{q_1}{q_2}+\frac{q_2}{q_1}\right)\mathcal{L}_1(\hat{q}_1 \cdot \hat{q}_1) \\
        &\quad + \frac{64}{245}\mathcal{L}_2(\hat{q}_1 \cdot \hat{q}_1)
        \bigg] f_{\nu}.
    \end{aligned}
    }
    \label{G2kernel}
\end{align}

{\bf These kernels are the work's main result, and are new; we refer to them as neutrino signature kernels.} From them, we see that neutrinos introduce a scale-dependent modification to the SPT kernels. 

This scale dependence can, in principle, produce significant effects on both the non-linear power spectrum via the $P_{22}$ and $P_{13}$ terms, and on the real- and redshift-space bispectrum. In real space, the bispectrum is influenced only by the $\mathcal{F}_2$ kernel, whereas in redshift space, it receives additional contributions from the $\mathcal{G}_2$ kernel and from the growth rate $f(k,\tau)$. As a result, we might expect the overall effect of the mode coupling of the neutrino signature kernels to be amplified in redshift space.

Let us examine the properties of the kernels we have obtained. One of the most important properties is that both \( \mathcal{F}_2(\mathbf{q}_1, \mathbf{q}_2) \) and \( \mathcal{G}_2(\mathbf{q}_1, \mathbf{q}_2) \) must scale as \( k^2 \) when \( \mathbf{k} \equiv \mathbf{q}_1 + \mathbf{q}_2 \) approaches zero, while \( \mathbf{q}_1 \) and \( \mathbf{q}_2 \) are not \citep{Bernardeau_2002}. This is consequence of momentum conservation in the center of mass frame \citep{Bernardeau_2002, szapudi1999n, goroff1986coupling}. By Taylor expanding the kernels (Eqs.~(\ref{F2kernel}) and (\ref{G2kernel})), we find that, at leading order in \( k \), the kernel corrections scale as \( k \) rather than \( k^2 \) and therefore, the momentum conservation is violated.

Let us also compare our results to a subsequent work \citep{Avilesfolpnu} following the initial submission of our paper. In an EdS Universe, their kernels are: 
\begin{align}
    \mathcal{F}_2(\mathbf{q}_1, \mathbf{q}_2) = F_2(\mathbf{q}_1, \mathbf{q}_2)
\end{align} 
and 
\begin{align}
    \mathcal{G}_2(\mathbf{q}_1, \mathbf{q}_2) &= G_2(\mathbf{q}_1, \mathbf{q}_2)- \bigg[
        \frac{91}{245} \mathcal{L}_0(\hat{q}_1 \cdot \hat{q}_1) \nonumber\\
        &\quad\qquad + \frac{3}{10}\left(\frac{q_1}{q_2}+\frac{q_2}{q_1}\right)\mathcal{L}_1(\hat{q}_1 \cdot \hat{q}_1)\nonumber \\
        &\quad\qquad + \frac{56}{245}\mathcal{L}_2(\hat{q}_1 \cdot \hat{q}_1)
        \bigg] f_{\nu}.
\end{align}
which are very close to our kernels in Eqs.~(\ref{F2kernel}) and (\ref{G2kernel}). The main difference is that the \( F_2 \) kernel (Eq.~(\ref{F2kernel})) receives an additional neutrino correction in our approach, whereas it does not receive any correction in \citep{Avilesfolpnu}. Furthermore, the coefficients of the \( G_2 \) kernel (Eq.~(\ref{G2kernel})) differ slightly between the two results. We conclude that although the assumptions between the two approaches are different, from the numerical perspective, we do not expect to see a significant difference between our kernels and \citep{Avilesfolpnu}. 


\section{\label{sec:level3} Redshift-Space Bispectrum and Neutrino Signature}

To maintain consistency, the kernels must be applied only to the higher-order density and velocity perturbations of the CDM+baryons since the neutrino density perturbations are kept at linear order and their velocity perturbations are zero.  Calculating the galaxy bispectrum from Eq (\ref{Eq:Bis_def}), using the bias relation from Eq.~(\ref{eqn:bias}) and the CDM+baryons density expansion from Eq.~(\ref{eqn:ansatz}), leads to terms such as \( \langle \delta_{\rm cb} \delta_{\rm cb} \rangle \), which correspond to the CDM+baryons auto-power spectrum, $P_{\rm cb}$, and terms like $\langle \delta_{\rm cb} \delta_{\nu} \rangle$, which produce the cross-power spectrum between neutrinos and CDM+baryons, $P_{\rm cb\nu}$. Therefore, the bispectrum is:
\begin{align}
    B(\mathbf{k_1}, \mathbf{k_2}, \mathbf{k_3}, \hat{n}) &= (1-3f_{\nu}) B_{\rm cb}(\mathbf{k_1}, \mathbf{k_2}, \mathbf{k_3}, \hat{n}) \nonumber\\&\quad+ 2f_{\nu}B_{\rm cb\nu}(\mathbf{k_1}, \mathbf{k_2}, \mathbf{k_3}, \hat{n})
    \label{RSD_B}.
\end{align}
We can calculate these power spectra directly from \textsc{class} \citep{CLASSCode}or use the approximate scheme of \citep{kamalinejad2022two}. The redshift-space bispectrum also depends on the line of sight, $\hat{n}$, which we often take to be in the $\hat{z}$ direction. Following \citep{Scoccimarro1999} we can write the expressions for $B_{\rm cb}$ and  $B_{\rm cb\nu}$ as:
\begin{align}
    B_{\rm cb}(\mathbf{k_1}, \mathbf{k_2}, \mathbf{k_3}, \hat{n}) &= 2 Z_1(\mathbf{k}_1) Z_1(\mathbf{k}_2)\\&\quad\times Z_2(\mathbf{k}_1, \mathbf{k}_2) P_{\rm cb}(k_1) P_{\rm cb}(k_2) + {\rm cyc.} \nonumber
    \label{eq:Bcb}
\end{align}
and 
\begin{align}
    B_{\rm cb\nu}(\mathbf{k_1}, \mathbf{k_2}, \mathbf{k_3}, \hat{n}) &=  2 Z_{1}(\mathbf{k}_1) Z_1(\mathbf{k}_2)\\&\quad\times Z_2(\mathbf{k}_1, \mathbf{k}_2) P_{\rm cb\nu}(k_1) P_{\rm cb}(k_2)\nonumber\\&\quad+2 Z_{1}(\mathbf{k}_1) Z_1(\mathbf{k}_3)\nonumber\\&\quad\times Z_2(\mathbf{k}_1, \mathbf{k}_3) P_{\rm cb\nu}(k_1) P_{\rm cb}(k_3) + {\rm cyc.}\nonumber
    \label{eq:Bcbv}
\end{align}
The $Z_1$ and $Z_2$ kernels are the redshift-space kernels and connect the \textit{galaxy density} perturbations to the \textit{total matter} density perturbation \citep{Scoccimarro1999}. These kernels depend on the SPT kernels $F_2$ and $G_2$, the growth rate, the galaxy biases, and the line of sight. The redshift-space kernels are written as \citep{Scoccimarro1999}:
\begin{align}
    Z_1(\mathbf{k}) &= b_1+f\mu^2,\nonumber\\
    Z_2(\mathbf{k}_1, \mathbf{k}_2) &= b_1 \mathcal{F}_2(\mathbf{k}_1, \mathbf{k}_2)+f\mu^2 \mathcal{G}_2(\mathbf{k}_1, \mathbf{k}_2)\\&\quad+\frac{f\mu k}{2}\left[\frac{\mu_1}{k_1}(b_1+f \mu_2^2)+\frac{\mu_2}{k_2}(b_1+f \mu_1^2)\right]\nonumber\\&\quad+\frac{b_2}{2}+ b_{\mathcal{G}_2}\left(\left(\frac{\mathbf{k}_1\cdot\mathbf{k}_2}{k_1 k_2}\right)^2-1\right),
\end{align}
where $\mu \equiv \hat{k}\cdot\hat{z}$ and $\mathbf{k}\equiv \mathbf{k}_1+\mathbf{k}_2$. The redshift-space bispectrum in Eq. (\ref{RSD_B}) depends on five parameters: the three triangle sides, $k_1$, $k_2$, and $k_3$; the cosine of the angle between $k_1$ and the line of sight, $\mu$; and the azimuthal angle of $k_2$ about $k_1$, which we denote by $\phi$, following \citep{scoccimarro1998nonlinear}. The cyclic summation in Eq. (\ref{eq:Bcb}) and Eq. (\ref{eq:Bcbv}) involves calculating $\mu_2$ and $\mu_3$, which are the cosines of the angles between $\mathbf{k}_2$ and $\mathbf{k}_3$ and the line of sight, respectively. They are:
\begin{align}
    x & \equiv \cos{\theta} = \frac{\mathbf{k}_1\cdot\mathbf{k}_2}{k_1k_2},\\
    \mu_2 &= \mu x - \sqrt{1-\mu^2}\sin{\theta}\cos{\phi},\\
    \mu_3 &= -\frac{k_1}{k_3}\mu-\frac{k_2}{k_3}\mu_2.
\end{align}
We have defined the $\theta$ as the angle between $k_1$ and $k_2$.

The bispectrum monopole is the average of Eq. (\ref{RSD_B}) over $\mu$ and $\phi$, as described in \citep{Scoccimarro1999}. 

We can explore the effects of the kernel updates presented in Eq. (\ref{F2kernel}) and (\ref{G2kernel}) by comparing the bispectrum monopole computed with and without the signature terms. We note that our calculations of the kernels are only valid on scales smaller than the free-streaming scale ($k>k_{\rm NR}$). To account for this, we apply a Heaviside function to the kernels and to the growth rate, $\Theta(k-k_{\rm NR})$, which affects only scales larger than $k_{\rm NR}$. 

Fig. \ref{B_ratio} shows the ratio of the bispectrum monopole including neutrino signature terms $B(k_1, k_2, k_3)$, to the bispectrum with the SPT kernels $\tilde{B}(k_1, k_2, k_3)$, for several neutrino masses. As the figure illustrates, the signature terms suppress the bispectrum monopole proportionally to the neutrino mass. The suppression  amplitude for a typical neutrino mass is less than 1\%.
\begin{figure}[H]
    {\includegraphics[width=0.45\textwidth]{Bispectrum_ratio_plot.pdf} }
    \caption{Ratios of the bispectrum with the signature terms in the kernels, $B(k_1, k_2, k_3)$, to the bispectrum with SPT kernels, $\tilde{B}(k_1, k_2, k_3)$, for three neutrino masses, largest in the top panel to smallest in the bottom panel. The suppression is scale-dependent and proportional to the neutrino mass, larger for larger masses. However, the suppression is less than 1\% for a typical neutrino mass. The green curve corresponds to $\sum m_{\nu} = 1;{\rm eV}$, the blue to $\sum m_{\nu} = 0.26;{\rm eV}$, and the red to $\sum m_{\nu} = 0.06;{\rm eV}$. This figure is produced by considering 372 closed triangles, with sides ranging from \( k_{\rm min} = 0.001\;[h/{\rm Mpc}] \) to \( k_{\rm max} = 0.5\;[h/{\rm Mpc}] \). Each wave number is binned in intervals of \( \Delta k = 0.035\;[h/{\rm Mpc}] \). This choice of binning will be used later in \S\ref{sec:synthetic}.}
    \label{B_ratio}
\end{figure}

\section{Synthetic Data}
\label{sec:synthetic}

In order to assess how the neutrino signature terms would impact parameter constraints from the bispectrum, we perform the following experiment. We create synthetic data that includes the signature contributions to the redshift-space bispectrum, mimicking the real bispectrum from galaxy surveys with error bars derived from the theoretical covariance matrix. 

We then fit two models to this data: one model employs the neutrino signature kernels (which should ideally return the original fiducial values), and the other uses the SPT kernels. This experiment aims to reveal to what extent  we may mismeasure the central value and the associated parameter covariance if the wrong kernels are used \citep{noriega2024unveiling}.

First, we explain why performing a traditional Fisher forecast would not be sufficient here. A Fisher forecast is not capable of obtaining the best-fit parameters. Our new kernels could affect the \textit{best-fit values} of the cosmological parameters as well as their variances. Therefore, in our case, a Fisher forecast would not provide the complete picture. 

In Appendix \S\ref{sec:MCMCvsFisher}, we compare the contour plots obtained from our MCMC analysis with those derived from a Fisher forecast. We demonstrate that there is no statistical difference between the two approaches. However, the MCMC method provides the additional advantage of yielding the best-fit parameter values for the given models. 

Our approach is statistically equivalent to generating numerous mock datasets and randomly shifting the bispectrum value on each triangle configuration by adding to it errors drawn from the covariance matrix. For instance, creating 10,000 mocks using this latter method, finding the best-fit parameters for each, and then averaging at the end, would be equivalent to averaging those mocks and then finding the best-fit parameters of the average. In this work, we adopt, essentially, the latter approach. The equivalence ultimately stems from the ergodic hypothesis.

\subsection{Mock Samples and Covariance}
To fully explore the potential of the neutrino signature, we perform the procedure explained above twice on two galaxy samples with different volumes. The first mock sample replicates the Sloan Sky Digital Survey (SDSS) Baryonic Acoustic Oscillation Spectroscopic Survey (BOSS) Data Release (DR) 12 CMASS galaxy sample \citep{reid2016sdss, SE_3PCF_BAO}, with an effective volume of $V_{\rm eff} = 3\;[{\rm Gpc}/h]^3$ which is comparable to that found by fitting a covariance matrix template to the 3PCF in \citep{SE_3PCF_BAO}. We also take the mean number density to be $\bar{n} \sim 3 \times 10^{-4}\;[h/{\rm Mpc}]^3$ at redshift $z = 0.57$ \citep{reid2016sdss}. The second sample simulates an upcoming survey with the same number density as BOSS CMASS DR12 but with a larger volume. Specifically, we consider a \textsc{DESI} Y5-like sample \citep{DESI:2016, grove2022desi}, with effective volume of $V_{\rm eff} = 25\;[{\rm Gpc}/h]^3$ at redshift $z=0.7$. We use a single tracer (Luminous Red Galaxies, LRGs)  with a constant number density $n(z)$ in all our analysis.

We then produce error bars from the Gaussian term of the covariance matrix ,$\mathbf{C}^{\rm PPP}$, as well as the contribution from the bispectrum, $\mathbf{C}^{\rm BB}$. We do not consider higher-order terms, namely, $\mathbf{C}^{\rm TP}$, and $\mathbf{C}^{\rm P6}$, which correspond respectively to trispectrum-power spectrum, and pentaspectrum contributions \citep{biagetti2022covariance}. The reason is that the Gaussian term dominates, and the off-diagonal contributions are much lower. These contributions should, in principle, be considered when one wants to carefully forecast the bispectrum on smaller scales. However, since these contributions are small, it is unlikely that they will significantly impact our result \citep{novell2024approximations}.

The Gaussian term is \citep{gualdi2020galaxy}:
\begin{align}
    \mathbf{C}^{\rm PPP} = &\frac{s_{123}\pi k_{\rm F}^3}{\Delta k^3}\frac{P(k_1) P(k_2)P(k_3)}{k_1 k_2 k_3 }b_1^6\\&\times\frac{1}{4\pi}\int_{0}^{2\pi}d\phi\int_{-1}^{1}d\mu\; \mathcal{P}(\mu, k_1)\mathcal{P}(\mu_2, k_2)\mathcal{P}(\mu_3, k_3),\nonumber
    \label{Cov}
\end{align}
where 
\begin{align}
    \mathcal{P}(\mu, k) = (1+\beta \mu^2)^2+\frac{1}{\bar{n}b_1^2P(k)},
\end{align} 
and the non-Gaussian term \citep{gualdi2019enhancing}:
\begin{align}
    \mathbf{C}^{\rm BB} = & \frac{k_{\rm F}^3\delta^{\rm K}_{34}}{4\pi k_3 k_4 \Delta k} \frac{1}{8}B(k_1, k_2, k_3)B(k_4,k_5,k_6)\nonumber\\&+{\rm 8\;perm.}\;,
    \label{Cov_BB}
\end{align}
where $\beta \equiv f/b_1$. We have used an approximation of $\mathbf{C}^{\rm BB}$ given by \citep{gualdi2019enhancing} rather than the exact form given by \citep{gualdi2020galaxy}. The Poisson noise (shot-noise) is produced because galaxies are discrete objects and not a continuous field, and its power spectrum is equal to one over the mean number density of galaxies, $\bar{n}$. The parameter $s_{123}$ encodes the shape of the triangle and is 6, 2, and 1 for respectively equilateral, isosceles, and scalene triangles.

We also note the the survey volume affects the overall amplitude of the covariance by changing $k_{\rm F}$, which is the fundamental mode of the survey. In a given survey with effective volume $V_{\rm eff}$, the smallest mode we can probe (the largest possible scale) has a wave-number of $k_{\rm F} = 2\pi/V_{\rm eff}^{1/3}$. In Eq. (\ref{Cov}), we have binned the Fourier space modes into intervals with a width of $\Delta k$. We fix $\Delta k = 0.035 \;h\;{\rm Mpc}^{-1}$ in all our analysis, to be consistent.

We note that Eq. (\ref{Cov}) assumes the thin-shell approximation, \textit{i.e.}, the spherical shells are thin relative to $k$. This approximation is valid on all scales except small $k$. However, these small $k$ are dominated by cosmic variance and therefore do not contribute much to the information content of the bispectrum and power spectrum. Thus, we fix the minimum $k$ in our analysis at $k_{\rm min} = 0.001\;[h/{\rm Mpc}]$ (which falls outside the range of validity for the thin-shell approximation but remains acceptable due to the reasoning previously discussed) and the maximum at $k_{\rm max} = 0.5\;[h/{\rm Mpc}]$. In Appendix \S\ref{sec:Binning}, we examine the effect of binning on the information extracted from the bispectrum. Given that the Fisher matrix accurately replicates the results obtained from an MCMC analysis, as demonstrated in Appendix \S\ref{sec:MCMCvsFisher}, we compute the Fisher matrices for two different bin widths. Our analysis shows no statistical difference between using $\Delta k = 0.035 \;h/{\rm Mpc}$ and $\Delta k = 0.02 \;h/{\rm Mpc}$, the latter being a more common choice (e.g., \citep{hahn2020constraining}). Consequently, we adopt $\Delta k$ as described.
\begin{table}[h!]
\centering
\label{tab:example}
\begin{tabular}{lcccr} 
\toprule
\textbf{Sample} & \textbf{$V_{\rm eff}\;[{\rm Gpc}/h]^3$} & \textbf{$\bar{n}\;[h/{\rm Mpc}]^3$} & \textbf{$\bar{z}$} & \textbf{Figure}\\ 
\midrule
CMASS DR 12 & $3$ & $3\times10^{-4}$ & $0.57$ & Fig. \ref{fig:mnu_260_V_3}\\ 
DESI Y5 & $25$ & $3\times10^{-4}$ & $0.7$ &Figs. \ref{fig:mnu_260_V_25} \& \ref{fig:mnu_1000_V_25}\\
\bottomrule
\end{tabular}
\caption{Summary of the survey characteristics considered in this paper. The first column is the sample name, the second the effective volume, the third the number density, the fourth the average redshift, and the fifth the Figure where the relevant results are displayed. Both samples represent Luminous Red Galaxies (LRGs).}
\end{table}
\subsection{Fiducial Cosmology \& Bispectrum Template}
We create  synthetic data from the bispectrum theoretical template from Eq. (\ref{RSD_B}) at a fiducial cosmology motivated by \textit{Planck} values \{$\sum m_{\nu} = 0.26\;{\rm eV}$, $\Omega_{\rm m} = 0.317$, $\Omega_{\rm b} = 0.049$, $\sigma_8 = 0.811$, $h = 0.67$, $n_{\rm s} = 0.96$ (spectral index of the primordial power spectrum)\}. We fix the baryons density $\Omega_{\rm b}$ to its \textit{Planck} best-fit values since LSS does not strongly constrain it.  We take as our fiducial galaxy bias parameters \{$b_1 = 2$, $b_2 = -1$, $b_{\mathcal{G}_2} = -0.35$\}.  $b_2$ is taken to be within $1\sigma$ of the values reported in \citep{ivanov2023cosmology, eggemeier2020testing} as their best fit. In our analysis, \( b_{\mathcal{G}_2} \) is set according to the local Lagrangian biasing, where $b_{\mathcal{G}_2} = -2/7(b_1 - 1)$,
as derived in \citep{Desjacques:2018}. We also keep the biases the same for all samples to be consistent.
\begin{table}[h!]
    \centering
    \caption{Summary of the fiducial cosmological parameters we used in creating the synthetic data.}
    \label{tab:summary_of_fid}
    \begin{tabular}{l c c c c c}
        \toprule
        Parameter & Value & Parameter & Value & Parameter & Value \\
        \midrule
        $b_1$ & 2 & $\sum m_{\nu}$ & 0.26 \& 1 eV & $h$ & 0.67 \\  
        $b_2$ & -1 & $\Omega_{\rm m}$ & 0.317 & $\sigma_8$ & 0.811 \\  
        $b_{\mathcal{G}_2}$ & -0.35 & $\Omega_{\rm b}$ & 0.049 & $n_{\rm s}$ & 0.96 \\   
        \bottomrule
    \end{tabular}
\end{table}
Since $b_1$ and $\sigma_8$ are highly correlated in the redshift-space bispectrum, we fix $\sigma_8$ to the value determined by \textit{Planck}. Additionally, our Fisher forecast (to be presented in future papers) indicates that the bispectrum does not strongly constrain $\sigma_8$. The error bar on $\sigma_8$ from \textit{Planck} is approximately $0.006$ \citep{aghanim2020planck}, while our forecast for a DESI Y5-like bispectrum sample suggests an error of about $0.033$, which is five times larger. This significant difference further justifies fixing $\sigma_8$ to its fiducial value.

\begin{figure}[H]
    {\includegraphics[width=0.47\textwidth]{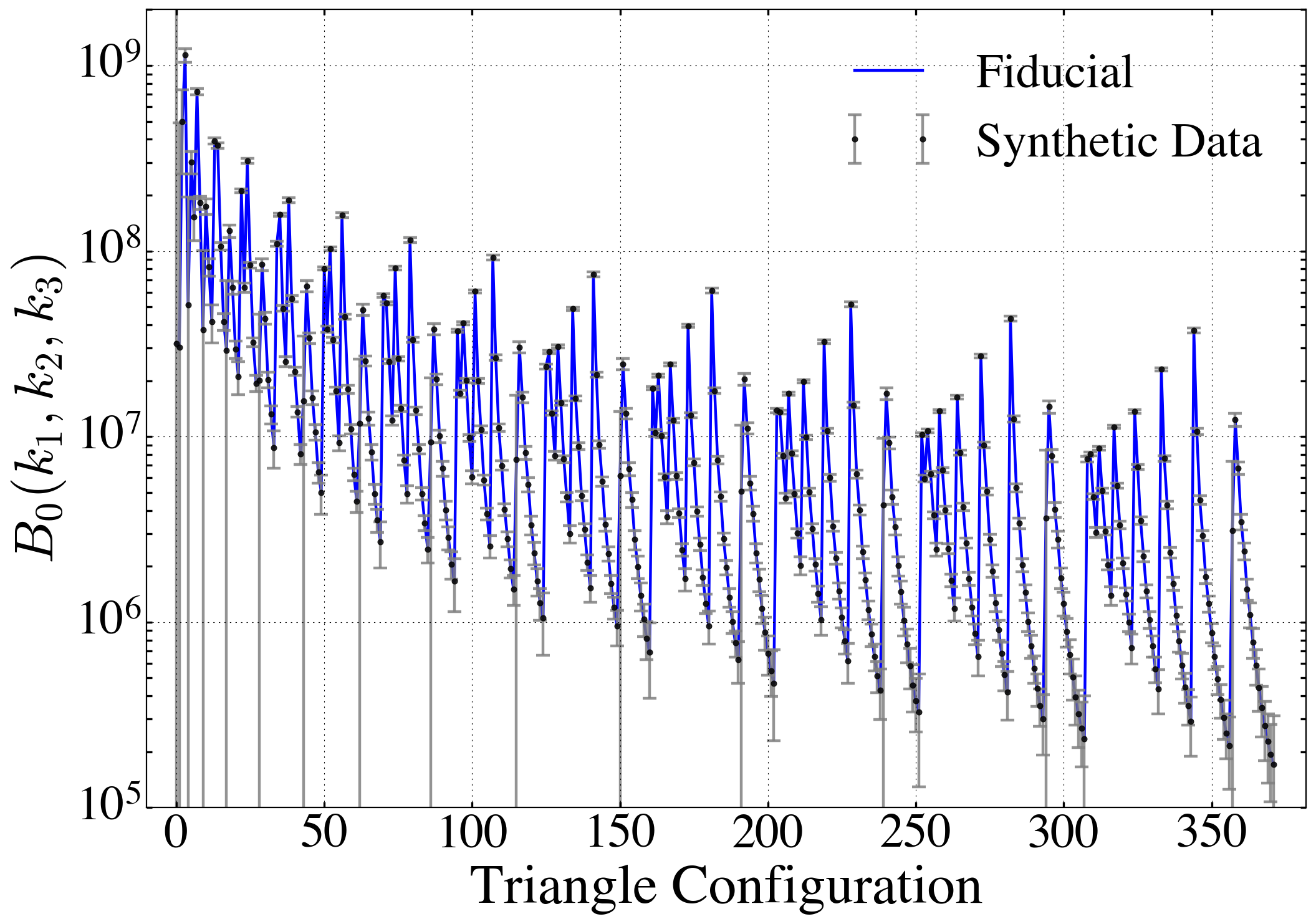} }
    \caption{The redshift-space bispectrum template with theoretical error bars for a BOSS CMASS-like survey at redshift of $\bar{z}=0.57$ and $\sum m_{\nu} = 0.26\;{\rm eV}$ as further explained in the main text (Table \ref{tab:summary_of_fid}). The triangle configurations have been obtained by considering \( k_1 \geq k_2 \geq k_3 \) and varying \( k_3 \), then \( k_2 \), and finally \( k_1 \).
}
    \label{syntheticdata}
\end{figure}

We also vary $\gamma = b_2/b_1$ and $\gamma^{'} = b_{\mathcal{G}_2}/b_1$ instead of $b_2$ and $b_{\mathcal{G}_2}$ themselves to reduce the  correlation with the linear bias. Therefore, our model contains four cosmological parameters, ($\sum m_{\nu}$, $\Omega_{\rm m}$, $h$ and $n_{\rm s}$) and three galaxy biases ($b_1$, $\gamma$ and $\gamma^{'}$). Fig. \ref{syntheticdata} shows the bispectrum synthetic data for all triangle configurations. Our binning leads to $372$ unique triangles with errorbars from Eq. (\ref{Cov}) and (\ref{Cov_BB}) obtained for BOSS CMASS DR12 survey volume and number density. Table \ref{tab:summary_of_fid} summarizes the fiducial cosmological parameters we considered in making our synthetic bispectrum data.

To more clearly show the signature term's impact, we also consider a heavier neutrino mass of $\sum m_{\nu} = 1\; {\rm eV}$ to increase the amplitude of the signature term. For this mass, the signature terms are no longer of order 1\%, but rather are about 4\%, according to Fig. \ref{B_ratio}. We emphasize that this neutrino mass is excluded by the BOSS power spectrum data at the $2\sigma$ level \citep{ivanov2020cosmological}. However, we consider it as an example to more strongly illustrate the effects of the kernels.

We now obtain the posterior distribution of the parameters, given the likelihood function. The likelihood function $\mathcal{L}$ depends on the data vector $\mathbf{D}$ (given the parameters, $\theta$), $\mathbf{M}$, which is either the bispectrum with our signature kernels or with the SPT kernels, and $\mathbf{C}^{-1}$, which is the inverse of the covariance matrix. We have
\begin{align}
    \ln{\mathcal{L}(\theta| \mathbf{D}}) = -\frac{1}{2}(\mathbf{D}-\mathbf{M})\mathbf{C}^{-1}(\mathbf{D}-\mathbf{M})^{\rm T}
    \label{Eq:log_likelihood}
\end{align}
To allow the MCMC to fully probe the parameter space, we use wide flat priors around the fiducial values to prevent skewing the chains in any particular direction. The chains are initialized by randomly sampling the parameters from the following ranges: $1<b_1 <5$, $-0.5<\gamma<1.5$, $-2<\gamma^{'}<1$, $0<\sum m_{\nu}<3.5\; {\rm eV}$, $0.25<\Omega_{\rm m}<0.45$, and $0.55<h<0.85\; \text{and}\;0.5<n_s<1.6$. These ranges also define the bounds for our priors. We have also ensured that the prior bounds are sufficiently wide, allowing the chains to remain well within the range of the priors. This allows the chains to freely explore the parameter space and converge to the best-fit value.

We use the \textsc{emcee} \textsc{Python} package for the MCMC implementation \citep{foreman2013emcee}, employing $55$ walkers (chains) for our parameter estimation. The random walks are terminated when either the number of steps reaches $10,000$ or the Gelman-Rubin (GR) statistic, $\hat{R}$, falls below $1.01$ \citep{gelman1992inference}. The Gelman-Rubin statistic compares the variance within each chain to the variance between chains. A value of $1.0$ indicates perfect convergence, while values significantly above this suggest that the chains have not yet mixed well and so the parameter estimates may be unreliable. We note that the walkers in the \textsc{emcee} package are not independent, as they interact with each other, which accelerates convergence. However, the GR statistic assumes independence of the walkers. Thus, in all of our MCMC analysis, we also empirically ensure convergence by visually tracking the chains, while also monitoring the GR statistic to further validate the convergence. We use the \textsc{GetDist} \citep{getdist} package for the contour plots presented later in the paper.
\subsection{Neutrino Mass Shift with Fixed Parameters}
As a simple test, let us fix all of the other parameters and vary only the neutrino mass to obtain the maximum likelihood. We first make a fiducial dataset, as explained previously, for the BOSS CMASS survey volume and with $\sum m_{\nu} = 0.06\;{\rm eV}$. We then find the log-likelihood function via Eq. (\ref{Eq:log_likelihood}). As we see from Fig. \ref{lnlikelihood}, the location of the best-fit shifts if we do not consider the signature terms. Using the SPT model shifts the best-fit value of the neutrino mass to \( \sum m_{\nu} = 0.033 \; {\rm eV} \), which, compared to the neutrino signature model, represents an almost $50\%$ change. We also obtain the error-bars on the neutrino mass from the Hessian matrix, which in this case is the second derivative of the log-likelihood function with respect to the neutrino mass. We find that the 95\% upper bound on \( \sum m_{\nu} \) changes from \( 0.248 \; {\rm eV} \) for the signature model to \( 0.157 \; {\rm eV}\) for the SPT model, an almost 37\% change. 
\begin{figure}[H]
    {\includegraphics[width=0.47\textwidth]{log_likelihood.pdf} }
    \caption{Shift of the minimum of the log-likelihood function (Eq. \ref{Eq:log_likelihood}) due to the signature kernels for a BOSS CMASS-like survey as summarized in Table (\ref{tab:example}) and \S\ref{sec:synthetic}. The blue curve is the log-likelihood of the bispectrum model with signature terms, while the red curve corresponds to the bispectrum with standard SPT kernels. The blue and red vertical dashed lines indicate the minima. This plot shows that neglecting the neutrino signatures shifts the central value toward smaller neutrino masses. We have kept all the parameters fixed except for the neutrino mass in this figure.}
    \label{lnlikelihood}
\end{figure}
The take-away from this simple test is that even when we know all parameters and fit for the neutrino mass, the signature terms induce a non-negligible effect, which is most relevant when working with N-body simulations, rather than real surveys. Ignoring the signature terms leads to the upper bound getting \textit{unrealistically} smaller. We also find a smaller central neutrino mass.

\section{Bispectrum Parameter Posterios}
\label{sec:Bis_Param_post}
\subsection{BOSS CMASS-like Sample }
\label{sec:BOSS}
In this section we present the results from our MCMC analysis of the redshift-space bispectrum. Fig. \ref{fig:mnu_260_V_3} shows the posterior distributions of the bispectrum models obtained from the MCMC chains for the BOSS CMASS volume with fiducial $\sum m_{\nu} = 0.26\;{\rm eV}$, which is at the 95\% confidence level upper bound of the \textit{Planck} best-fit. We first notice that the central values of the contours are not the same as the fiducial for both of the models. The median values of the chains are about half a sigma away from the fiducial cosmological parameters. We anticipate that this is due to the small volume of BOSS CMASS sample since we do not observe this effect for DESI Y5 volume (which will be discussed later).

We also notice that there is a `tail' on the neutrino mass ellipses. Since the error-bars on the bispectrum are large, the MCMC allows for larger neutrino mass. Since the volume probed is small, the neutrino mass is not constrained very well, causing a tail on other parameters as well (for example, $\Omega_{\rm m}$, $n_{\rm s}$ and $h$). This tail occurs because an increase in the neutrino mass mean more suppression of the power spectrum, necessitating a compensatory increase in the matter density $\Omega_{\rm m}$ or the spectral index $n_{\rm s}$. Due to the limited constraining power of the BOSS CMASS volume on neutrino mass, it permits significantly higher masses, which, in turn, influence the estimates of $\Omega_{\rm m}$, $n_{\rm s}$ and $h$. 

We also observe that the best-fit linear bias has shifts by nearly half a sigma down when using the SPT models. At $2\sigma$, the linear bias is significantly influenced by neglecting the signature kernels. This is again due to large error-bars on the neutrino mass, which allows for larger $n_{\rm s}$ and $\Omega_{\rm m}$, which in turn allows $b_1$ to have smaller values since $b_1$ has negative correlation with $\Omega_{\rm m}$.

To mitigate the skewing of the chains towards higher neutrino masses at $2\sigma$, one approach is to impose tighter priors on the cosmological parameters. In principle, we could apply Gaussian priors around the parameters to prevent this tail, but we opted not to add additional information to the parameters, as this could impact the recovery of best-fit values.

Fig. \ref{fig:mnu_260_V_3} shows that the BOSS survey volume is not sufficient when it comes to precise measurement of the neutrino mass. This is also shown in the previous work done by \citep{ivanov2020cosmological} where they conclude that the BOSS power spectrum data can only be used to break degeneracies within the \textit{Planck} data, rather than probing the free-streaming effects. Many previous studies have shown that using the bispectrum can tighten the constraints on the cosmological parameters (for instance \citep{bispectrumneutrinolensing, yankelevich2023halo, Hahn_2021, hahn2020constraining}). \citep{d2024boss} fits the bispectrum and power spectrum at 1-loop to the BOSS data and finds that including the bispectrum will remarkably reduce the error-bars on the parameters, finding a 30\% reduction in the error-bar of $\sigma_8$ over the power spectrum. However, they fix the neutrino mass in their analysis (similar to the analysis done by \citep{philcox2022boss}). 
\begin{figure*}[!htb]
    \centering
    \includegraphics[width=\textwidth]{MCMC_kmax50_mnu_260_V_3_z_57.pdf}
    \caption{Posterior distribution obtained from MCMC on the bispectrum synthetic data described in \S\ref{sec:synthetic}. The volume used in this forecast (summarized in Table \ref{tab:example}) is similar to match BOSS CMASS, $V_{\rm eff} = 3\;[{\rm Gpc}/h]]^3$ with $\sum m_{\nu} = 0.26\;{\rm eV}$ at redshift $\bar{z}=0.57$. The red and black contours show the $1\sigma$ and $2\sigma$ confidence intervals corresponding to the bispectrum with neutrino signature kernels (Eq. (\ref{F2kernel}) and Eq. (\ref{G2kernel})) and with SPT kernels (Eq. (\ref{Eq:F2}) and Eq. (\ref{Eq:G2})). ($\times$) shows the central value of the SPT model while the horizontal and vertical dashed lines represent the fiducial values. At $2\sigma$, the linear bias is shifted notably while the higher-order biases and cosmological parameters are not affected. Since the sample volume is small, larger neutrino masses are allowed, leading to a tail in the posterior distribution. This results in corresponding tails in the cosmological parameter distributions, compensating for the larger suppression of the power spectrum due to the increased neutrino mass.}
    \label{fig:mnu_260_V_3}
\end{figure*}


\subsection{DESI Y5 Volume}
\subsubsection{$\sum m_{\nu} = 0.26\;{\rm eV}$}
\label{sec:DESI I}
We now increase the volume to $V_{\rm eff} = 25\;[{\rm Gpc}/h]^3$, and with the same number density as before, but at a higher redshift of $\bar{z} = 0.7$. The fiducial neutrino mass is $\sum m_{\nu} = 0.26\;{\rm eV}$ (used in BOSS CMASS we discussed in \S\ref{sec:BOSS}).
Figure \ref{fig:mnu_260_V_25} shows the contour plots resulting from the MCMC. We first notice that the neutrino mass ($\sum m_{\nu}$) is consistent with zero at $2\sigma$. The parameter constraints improve compared to BOSS CMASS (Fig. \ref{fig:mnu_260_V_3}). The tail that we observed in the BOSS CMASS sample disappears since the volume is much larger (about 8 times the CMASS volume). The SPT kernels have caused a shift in the best-fit value of the linear bias $b_1$ and  the tidal tensor bias, $b_{\mathcal{G}_2}$, with almost an $1\sigma$ shift for $b_1$. It appears that the other parameters are not affected by the kernels at all.  We suspect that this lack of effect is because the constraints on the other cosmological parameters come predominantly from the linear power spectrum in Eq. (\ref{eq:Bcb}) and Eq. (\ref{eq:Bcbv}). Therefore, a sub-percent change due to the extra terms induced by our signature kernels does not affect the estimation of cosmological parameters.

Another notable feature in Fig. \ref{fig:mnu_260_V_25} is that the biases exhibit minimal degeneracy with the cosmological parameters, except for the correlation observed between the linear bias $b_1$ and these parameters. Given that the impact of neutrinos can be effectively modeled as an overall suppression (including their impact on the kernels as shown in Fig. \ref{B_ratio}), the linear bias $b_1$ is primarily influenced.

There are several studies that provide insight on neutrino mass measurements from future galaxy surveys. \citep{RaccoZhangHenry} shows that using \textit{Planck}+DESI power spectrum and bispectrum will shrink the $1\sigma$ error-bar on the neutrino mass to  $15\;{\rm meV}$. DESI Y1 full-shape power spectrum provides 95\% upper bound of $\sum m_{\nu}<0.409\;{\rm eV}$ which is independent of the CMB \citep{DESIFullshape}. Other forecasts of future LSS+CMB experiments such as \citep{carbone2011neutrino, di2014cosmological, brinckmann2019promising, chudaykin2019measuring, archidiacono2025euclid} reveal that the 95\% upper bound on the neutrino mass will be less than $100\;{\rm eV}$, shedding light on the mass ordering.
\begin{figure*}[!htb]
    {\includegraphics[width=\textwidth]{MCMC_kmax50_mnu_260_V_25_z_7.pdf} }
    \caption{Same as Fig. \ref{fig:mnu_260_V_3} but with $V_{\rm eff} = 25\;[{\rm Gpc}/h]^3$, $\sum m_{\nu} = 0.26\;{\rm eV}$ at redshift $\bar{z}=0.7$ corresponding to DESI Y5 sample (refer to Table \ref{tab:example}). The linear bias estimation has shifted by about $1\sigma$ if the signature kernels are not used. There is also a small shift in the tidal tensor bias.}
    \label{fig:mnu_260_V_25}
\end{figure*}

\subsubsection{$\sum m_{\nu} = 1\;{\rm eV}$}
We now increase increase the neutrino mass to an unrealistic fiducial value of $\sum m_{\nu} = 1\;{\rm eV}$, with the survey specifications kept the same as in \S\ref{sec:DESI I} (DESI Y5-like). Fig. \ref{fig:mnu_1000_V_25} shows the contour plots from the MCMC. Similarly to what we found in \S\ref{sec:DESI I}, the best-fit values of the biases shifts significantly if we use the SPT kernels. The cosmological parameters we recover are not affected by the kernels since the power spectrum is the dominant source of the constraints on them.
\begin{figure*}[!htb]
    {\includegraphics[width=\textwidth]{MCMC_kmax50_mnu_1000_V_25_z_7.pdf} }
    \caption{Same as in Fig. \ref{fig:mnu_260_V_3} but with $V_{\rm eff} = 25\;[{\rm Gpc}/h]^3$ and $\sum m_{\nu} = 1\;{\rm eV}$ at redshift $\bar{z}=0.7$.  The linear and tidal tensor biases we recover have shifted by about $2\sigma$ from their inputted values.}
    \label{fig:mnu_1000_V_25}
\end{figure*}

\section{Conclusion}
In this paper, we obtained PT kernels including the neutrino mass and determined the redshift-space bispectrum monopole with respect to the line of sight. Using the theoretical covariance matrix, we fitted the bispectrum model with both the SPT kernels and the neutrino signature kernels to synthetic data generated from the signature model. We found that neglecting the signature terms in the bispectrum leads to an underestimation of the neutrino mass (Fig. \ref{lnlikelihood}) when all other parameters are known (as can occur with N-body simulations, where the ground truth is already available). In a real fitting procedure, this information is not available, and an MCMC is required to find the best-fit values.

We found that using the SPT kernels results in a shift in the recovered parameters, which is more pronounced for the galaxy biases. The cosmological parameters are not significantly affected because most of the information comes from the power spectrum rather than the kernels. Thus, a sub-percent change in the bispectrum does not substantially impact the cosmological parameters. However, this is not the case for the bias parameters. For a DESI Y5-like survey, we found that the kernels shift the bias parameters by almost $1\sigma$.

This work shows that correct modeling of the neutrino effects on the bispectrum is crucial when it comes to obtaining the galaxy biases. In our future work, we will explore how the kernels can impact the information obtained in a joint analysis of the power spectrum  and bispectrum. 

\begin{acknowledgments}
The authors thank Giovanni Cabass, Daniel Eisenstein, Simone Ferraro, ChangHoon Hahn, Misha Ivanov, Alex Krolewski, Stephen Portillo, and Matias Zaldarriaga for useful conversations. We particularly thank Alex Krolewski for comments on the draft and on adding in cosmological parameter errors, Simon Foreman for extremely extensive comments on the draft and discussion, and Matias Garny and Petter Taule for enlightening discussion and performing and sharing some unpublished calculations. We also thank Alejandro Aviles for useful comments and discussions. We thank all Slepian group members for helpful conversations over the years, especially Jiamin Hou and Alessandro Greco. ZS acknowledges funding from NASA grant number 80NSSC24M0021.
\end{acknowledgments}

\appendix
\section{Comparison of MCMC with Fisher Forecast}
\label{sec:MCMCvsFisher}
It is worth assessing how the MCMC approach employed in this paper compares to a traditional Fisher forecast. To do so, we first calculate the Fisher matrix for the DESI Y5 volume and number density, as outlined in the main text \S\ref{sec:synthetic}, using the fiducial cosmology previously specified (set by \textit{Planck}: \{$\sum m_{\nu} = 0.26\;{\rm eV}$, $\Omega_{\rm m} = 0.317$, $\Omega_{\rm b} = 0.049$, $\sigma_8 = 0.811$, $h = 0.67$, $n_{\rm s} = 0.96$). The Fisher matrix elements are:
\begin{align}
F_{ij} \equiv \frac{\partial B}{\partial \theta_i}\mathbf{C}^{-1}\frac{\partial B}{\partial \theta_j}
\label{Eq:Fisher}
\end{align}
where $B$ denotes the bispectrum monopole, $\mathbf{C}^{-1}$ is the inverse of the covariance matrix, $\theta$ represents the parameters of our model, and the indices $i$ and $j$ indicate the element of the matrix.

Figure \ref{fig:MCMCvsFisher} shows the comparison between the $1\sigma$ and $2\sigma$ error-bars obtained from the MCMC and Fisher forecast. Blue represents the error-bars of obtained from the MCMC and red are the error-bars obtained from Fisher forecast. As we can see, the error-bars are almost identical, with very little difference between them, which is not significant at all. The MCMC has over-estimated the error-bars by a small fraction of the standard deviation, which is more noticeable for $b_2$, $b_{\mathcal{G}_2}$ and $\sum m_{\nu}$. The correlation between the parameters are similar in both approaches as well. Therefore, this method gives the same result as the Fisher information would, as expected.

\begin{figure*}[!htb]
    {\includegraphics[width=\textwidth]{MCMC_Vs_Fisher.pdf} }
    \caption{Comparison of the $1\sigma$ and $2\sigma$ error bars on the parameters from the bispectrum monopole for DESI Y5 survey characteristics \ref{tab:example}, obtained using the Fisher matrix (red) and the MCMC analysis described in the main text (blue). The error bars and parameter correlations are nearly identical, with only minor differences that are not statistically significant. This demonstrates that our approach produces results consistent with a Fisher forecast while additionally providing the best-fit parameter values. In this paper, we have shown that the use of SPT kernels induces shifts in the recovered best-fit parameter values, which prevents the use of a Fisher forecast alone.}
    \label{fig:MCMCvsFisher}
\end{figure*}
\section{Binning Width Effect on the Posterior Distribution}
\label{sec:Binning}
In this appendix, we give a more detailed justification for our choice of binning width in the analysis. Specifically, in our MCMC study, we use a bin width of $\Delta k = 0.035\; h/\mathrm{Mpc}$. Here, we compare that to an alternative bin width of $\Delta k = 0.02\; h/\mathrm{Mpc}$. The motivation for this comparison is twofold: first, the narrower binning allows for the inclusion of more triangles. Second, it provides finer bins that could potentially capture scale-dependent effects more effectively.

Figure \ref{fig:binning} presents the corner plot of the posterior distribution of the redshift-space bispectrum monopole (Eq. \ref{RSD_B}) obtained using a Fisher forecast framework. We calculate the Fisher information matrix using Eq. (\ref{Eq:Fisher}). The analysis employs the fiducial cosmology described in \S\ref{sec:synthetic} of the main text for the DESI Y5 volume and number density. The only variable modified in this comparison is the binning width, $\Delta k$.

The corner plot reveals no significant differences between the two binning choices. Employing the finer bins ($\Delta k = 0.02\; h/\mathrm{Mpc}$) results in only a slight increase in the error bars for $h$ and $n_{\mathrm{s}}$, which is not statistically significant. Furthermore, the parameter correlations remain unchanged between the two binning schemes.

We interpret this result as an empirical validation of our original choice of bin width, as discussed in the main text. This robustness demonstrates that the adopted binning scheme adequately balances the trade-offs between computational efficiency and capturing relevant scale dependencies.
\begin{figure*}[!htb]
    {\includegraphics[width=\textwidth]{contour_plot_bin_comparison.pdf} }
    \caption{Comparison of the $1\sigma$ and $2\sigma$ error bars on the parameters obtained from Fisher forecasting of the bispectrum monopole for DESI Y5 survey characteristics, using two different binning widths. The blue contours correspond to $\Delta k = 0.02 \;h/{\rm Mpc}$, while the red contours represent $\Delta k = 0.035 \;h/{\rm Mpc}$. Fig. \ref{fig:MCMCvsFisher} shows that the predictions of the MCMC and Fisher forecast are statistically the same, so here we just use the Fisher forecast approach. Choosing finer bins increases the number of triangles, making the computation more expensive. However, as shown in the plot, the error bars do not significantly change with varying bin widths. This is likely because the neutrino effects in $k$-space manifest as smooth, broad suppression rather than sharp features. Consequently, this plot supports the binning choice adopted in the main text.}
    \label{fig:binning}
\end{figure*}

\bibliographystyle{apsrev4-1}
\bibliography{non-degenerate_neutrino_signature}
\end{document}